\documentclass[prl,aps,showpacs,twocolumn]{revtex4}
\usepackage{graphicx}
\usepackage{amsmath}
\usepackage{amsfonts}
\usepackage{amssymb}

\begin{document}

\title{Motor driven microtubule shape fluctuations - force from within the lattice}
\author{Herv\'{e} Mohrbach$^{1}$ and Igor M. Kuli\'{c}$^{2}$\\$^{1}$Laboratoire de Physique Mol\'{e}culaire et des Collisions,
Universit\'{e} Paul Verlaine - 57012 Metz, France\\$^{2}$School of Engineering and Applied Sciences, Harvard University,
Massachusetts 02138, USA}
\date{\today}
\begin{abstract}
We develop a general theory of microtubule (MT) deformations by molecular
motors generating internal force doublets within the MT\ lattice. We describe
two basic internal excitations, the S and V shape, and compare them with
experimental observations from literature. We explain the special role of
tubulin vacancies and the dramatic deformation amplifying effect observed for
katanin acting at positions of defects. Experimentally observed shapes are
used to determine the ratio of MT shear and stretch moduli ($\approx
6\times10^{-5}$) and to estimate the forces induced in the MT lattice by
katanin (10's of pN). For many motors acting on a single MT we derive
expressions for the end-to-end distance reduction and provide criteria for
dominance of this new effect over thermal fluctuations. We conclude that
molecular motors if acting cooperatively can ''animate'' MTs from within the
lattice and induce slack even without cross-bridging to other structures, a
scenario very much reminiscent of the motor driven axoneme.
\end{abstract}


\maketitle

Microtubules are the stiffest cytoskelletal component and constitute the main
routes for motor mediated intracellular cargo transport in higher organisms
\cite{Kinesin/Dynein}. Understanding their physical properties is at the heart
of many biological problems from cellular mechanics to information and
material trafficking in the cell. Since the discovery of their high elastic
anisotropy \cite{Kis} it became increasingly clear that MTs are mechanically
more complex than other semiflexible biofilaments. The high anisotropy has
been impressively confirmed by thermal fluctuation analysis of beads attached
to MTs of different lengths\cite{Pampaloni}. The emerging picture of the
MT\ is that of an anisotropic fiber reinforced material \cite{Kis,Pampaloni}
with the tubulin protofilaments (PF) acting as strong fibers weakly linked
with easily shearable lateral bonds. Remarkably this type of design is also
found in higher structures like axonemes (constituting the backbone of
flagella and cilia) where relatively inextensible MTs are held together with
highly stretchable nexin connections \cite{Kinesin/Dynein}. This remarkable
structural self-similarity of the two nested structures (MT and axoneme)
indicates further analogies in the way they respond to external and internal
forces. We explore here important consequences of MT geometry and elastic
properties and show that motors acting on the MT\ surface can generate
internal lattice strains sufficient to induce observable lateral and
longitudinal deformations of the MT\ backbone.

In the following we describe a twist-free MT of length $L$ consisting of $N$
identical PFa with constant distance $a$ and a circular cross section, Fig 1.
Each PF, parametrized by the MT backbone arc length $s$ has a position
dependent displacement $u_{k}\left(  s\right)  $ from its equilibrium
position, $k=1,...N$. The backbone shape is described by a curvature vector
$\vec{\kappa}\left(  s\right)  =\frac{d}{ds}\vec{t}\left(  s\right)  $ with
$\vec{t}\left(  s\right)  $ the bundle centerline tangent. The elastic
properties of the MT are characterized by a PF bending stiffness $B=\frac
{1}{64}\pi a^{4}Y$ and compressional modulus $K_{c}=\frac{\pi}{4}a^{2}%
Y\ $\ with $Y\approx0.1-1.5GPa$ \ \cite{Kis,Pampaloni} being the PF Young's
modulus. Additionally there are shear elastic forces restoring the
longitudinal displacement between the PFs governed by a very soft elastic
shear modulus $K_{s}\approx10^{-3}-1MPa$ \cite{Kis,Pampaloni}$.$ The elastic
energy is given by
\begin{equation}
E_{MT}=\frac{1}{2}\sum\nolimits_{k=1}^{k=N}\int\nolimits_{-L/2}^{L/2}\left(
B\vec{\kappa}^{2}+K_{c}u_{k}^{\prime2}+K_{s}\Delta_{k}^{2}\right)  ds
\label{Etot_def}%
\end{equation}
with the first term being the bending energy, the second the PF compression
and the third describing the relative shear energy between the neighboring
PFs. The shear displacement $\Delta_{k}$ is related to the difference of PF
displacements $u_{k}-u_{k-1}$ and a curvature induced additional displacement
via
\begin{equation}
\Delta_{k}\left(  s\right)  =u_{k}\left(  s\right)  -u_{k-1}\left(  s\right)
+\int\nolimits_{-L/2}^{s}\vec{\kappa}\left(  s^{\prime}\right)  \cdot
\Delta\vec{r}_{k}ds^{\prime} \label{Delta_k}%
\end{equation}
With $\Delta\vec{r}_{k}=\vec{r}_{k}-\vec{r}_{k-1}$ and $\vec{r}_{k}%
=R_{MT}\left(  \cos\frac{2\pi k}{N},\sin\frac{2\pi k}{N}\right)  $ the vector
pointing from the MT center to the k-th PF, cf. Fig 1. Equations
\ref{Etot_def}-\ref{Delta_k} are 3-D analogues of the previously proposed
stretchable railway-track \cite{Everaers} or wormlike-bundle \cite{Heussinger}
model for the case of a hollow circular bundle. While in general all the $N+3$
fields , i.e. the 3 components of $\vec{\kappa}\left(  s\right)  $ and the PF
displacements $\left\{  u_{k}\left(  s\right)  \right\}  _{k=1,..,N}$ enter
the eqs \ref{Etot_def}-\ref{Delta_k} in the limit of small MT deviations from
a straight line the problem can be drastically simplified. We first expand the
tangent $\vec{t}\approx\left(  \theta_{x},\theta_{y},1\right)  $ and
$\vec{\kappa}\approx\left(  \theta_{x}^{\prime},\theta_{y}^{^{\prime}%
},0\right)  $ in terms of two angular projections $\theta_{x}$ and $\theta
_{y}$ of $\vec{t}$ in $x$ and $y$ direction respectively. Exploiting the
circular geometry of the PF arrangement and the Fourier representation
$u_{k}\left(  s\right)  =\sum\hat{u}_{q}\left(  s\right)  e^{\frac{2\pi
ikq}{N}}$ over $k$ we quickly realize that only the longest wavelength mode
$\hat{u}_{1}\left(  s\right)  $ couples to overall MT backbone shape given by
the curvature $\vec{\kappa}$. This leads to total energy decoupling
$E_{MT}=E_{MT}^{0}+E_{MT}^{x}+E_{MT}^{y}$ into a shape- independent component
$E_{MT}^{0}$ and two shape dependent contributions (in x and y direction)
given by:
\begin{equation}
E_{MT}^{i}=\frac{1}{2}\int\nolimits_{-L/2}^{L/2}\left[  \hat{B}\Delta
\theta_{i}^{\prime2}+\hat{K}_{c}U_{i}^{\prime}{}^{2}+\hat{K}_{s}\hat{\Delta
}_{i}^{2}\right]  ds\text{ }%
\end{equation}
With $\hat{\Delta}_{i}\left(  s\right)  =a\left(  \theta_{i}\left(  s\right)
-\theta_{i}\left(  -L/2\right)  \right)  -U_{i}\left(  s\right)  $ ,
$a=\left|  \Delta\vec{r}_{k}\right|  $ the inter-protofilament distance ,
$i=x,y$ and $\vec{U}\left(  s\right)  =\left(  U_{x},U_{y}\right)  =\left(
\text{Re}\chi\hat{u}_{1},\text{Im}\chi\hat{u}_{1}\right)  $ with
$\chi=1-e^{-2\pi i/N}$ and renormalized constants $\hat{B}=NB,$ $\hat{K}%
_{c}=NK_{c}/\left(  4-4\cos\left(  2\pi/N\right)  \right)  $ and $\hat{K}%
_{s}=NK_{s}$. Visually the new variable $\vec{U}\left(  s\right)  $ is a x-y
vector at each MT\ -crosscut and can be interpreted as the (vectorial) mean
over relative PF displacements of neighboring PFs. With this enormous
simplification at hand we can consider now basic motor induced MT excitations
(Fig 1). There are two elementary configurations in which motors can induce
internal MT strains: 1) A motor (or a complex of several motors) acting
between two (not necessarily neighboring) PFs and 2)\ A motor (or a complex of
several motors) acting at two points within the same PF. For reasons that will
soon become clear we call the excitation 1 S-type or simply an
\textit{''S-let''} and excitation 2 we call an V-type excitation or
\textit{''V-let''}. Both excitations are ''internal'' in the sense that there
is no net torque or force on the system motor+MT similarly to the case of a
beating flagellum\cite{Kinesin/Dynein}.

\begin{figure}[ptb]
\includegraphics*[width=8cm]{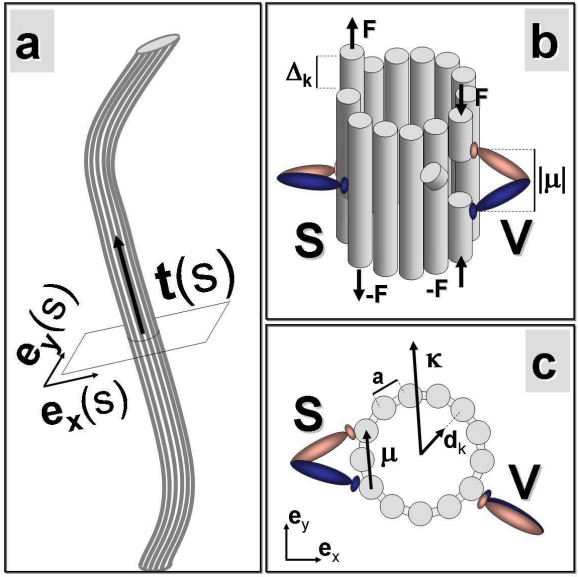}\caption{(Color online) The basic
geometry of motors inducing internal force doublets along the MT backbone:
between two PFs (S-let) and along the same PF (V-let). The red and blue ovals
represent two coupled motors or two motor subunits (legs) of the same motor.
Tubulin lattice vacancies at the motor position strongly amplify the MT
backbone deformation ($V^{gap}$ -let). }%
\label{KinkCompilation1}%
\end{figure}

\textit{Elementary internal MT excitations.} In the following we want to
understand the properties of the two basic types of excitations from Fig 1 and
focus on the S-type first. We assume a single motor (or a complex of two
motors) at position $s=s_{0}$ bridging between two PF with index $k_{1}$ and
$k_{2}$ and exerting opposing forces $F$ and $-F$ onto them respectively, Fig
1 a) (left) + b). The total energy is $E_{tot}=E_{MT}+E_{S-mot}$ with $E_{MT}$
given by eqs \ref{Etot_def}-\ref{Delta_k} and the potential energy of the
motor $E_{S-mot}=-F\sum_{k=k_{1}}^{k=k_{2}}\int_{-L/2}^{L/2}\delta\left(
s-s_{0}\right)  \Delta_{k}\left(  s\right)  ds$ . As we had for $E_{MT}$
before $E_{S-mot}$ also decouples into independent modes in the Fourier
representation over $k$ and $E_{mot,S}=E_{S-mot}^{0}+$ $E_{S-mot}%
^{x}+E_{S-mot}^{y}$ with $E_{S-mot}^{0}$ a curvature independent term and the
two shape dependent contributions $E_{S-mot}^{x/y}$ (in x and y direction)
given by: $E_{S-mot}^{i}=-F\frac{\mu_{i}}{a}%
{\textstyle\int\nolimits_{-L/2}^{L/2}}
\delta\left(  s-s_{0}\right)  \hat{\Delta}_{i}\left(  s\right)  ds$ with
$i=x,y$ and $\vec{\mu}$ being the vector connecting the two attachment points
of the motor (or motor complex) with components $\mu_{i}=\vec{\mu}\cdot\vec
{e}_{i}$ , Fig 1 b. The equilibrium solution is given by the Euler Lagrange
equations: $\delta E_{tot}^{i}/\delta U_{i}=0$ and $\delta E_{tot}^{i}%
/\delta\theta_{i}=0$ with $E_{tot}^{i}=E_{S,mot}^{i}+E_{MT}^{i}$ and boundary
conditions $\theta_{i}^{\prime}\left(  \pm L/2\right)  =U_{i}^{\prime}\left(
\pm L/2\right)  =0$ (vanishing bending and shearing stresses at the ends). A
short calculation leads to $\hat{\Delta}_{i}^{\prime\prime}-\lambda^{-2}%
\hat{\Delta}_{i}=\left(  1+\alpha\right)  \frac{\mu_{i}F}{a\hat{K}_{c}}%
\delta\left(  s-s_{0}\right)  $ with the shear decay length $\lambda=\left(
\lambda_{c}^{-2}+\lambda_{B}^{-2}\right)  ^{-1/2}$. Here the two important
length scales $\lambda_{c}=\sqrt{\hat{K}_{c}/\hat{K}_{s}}$ and $\lambda
_{B}=\sqrt{\hat{B}/a^{2}\hat{K}_{s}}$have the physical meaning of a pure
compression- / pure bending- induced shear screening length respectively, with
their squared ratio $\alpha=\left(  \lambda_{c}/\lambda_{B}\right)
^{2}\approx35$ ($N=13$ PF). The remaining equations lead to conservation laws
$\left(  a\theta_{i}+\alpha U_{i}\right)  ^{\prime}=0$ which combined with the
equation for $\hat{\Delta}_{i}$ give for the simplest symmetric case $s_{0}=0$
the tangent angles (up to an arbitrary constant)
\begin{equation}
\theta_{i}\left(  s\right)  =\Phi_{i}\frac{\cosh\left(  \left|  s\right|
/\lambda-L/2\lambda\right)  }{\sinh\left(  L/2\lambda\right)  }\text{
}\label{thetaSType}%
\end{equation}
With $\Phi_{i}=\Phi_{i}^{S}=\vec{\mu}\cdot\vec{e}_{i}\tfrac{\lambda F}%
{2\hat{B}}.$ The resulting MT backbone curvature has a jump at $s=0$ and
attains its maximal modulus there. The resulting MT shape is planar (contained
in the plane spanned by $\vec{\mu}$ and $\vec{t}$ at any position) and
S-shape-like with initial and final angle coinciding $\theta_{i}\left(
-L/2\right)  =\theta_{i}\left(  +L/2\right)  $ which explains our nomenclature
''S-type excitation'' or ''S-let''. Interestingly the length scale $\lambda$
over which $\theta_{i}\left(  s\right)  $ declines allows us to independently
estimate the ratio of stretch and shear moduli from the observation of S-let
deformations coming from katanin action\cite{MTSTypeKink}, $Y/K_{s}\approx$
$64\pi^{-1}\left(  \lambda/a\right)  ^{2}\approx6\times10^{5}$ for
$\lambda\approx1\mu m$ (cf, Fig 2b) and $a\approx6nm.$ This value is close to
the result obtained by Pampaloni et al. \cite{Pampaloni} ($Y/K_{s}%
\approx10^{6}$). Further the maximal deflection angle of $\theta_{\max}%
\approx38^{\circ}$ ($=0.66$) measured in Fig 2a gives via \ref{thetaSType} an
estimate for the involved motor forces $F\approx2\hat{B}\theta_{\max}%
/(\lambda\left|  \mu\right|  )=\frac{13}{32}\pi a^{4}Y/(\lambda\left|
\mu\right|  )=20-250pN$ (for $\left|  \mu\right|  =a\approx6nm$ and the range
of values for $Y$ from literature \cite{Kis,Pampaloni}). This indicates that
many katanin motors might act cooperatively to generate the observed shape
change in Fig 2b. Another interesting possibility is that katanin might be
different from dynein/kinesin by generating only small contractile
displacement ''powerstrokes'' $\lesssim1nm$ with a more efficient
chemical-mechanical ATP-energy conversion leading to larger contractile forces
$F$ $\gtrsim15k_{B}T/nm=60$ $pN$.

The second fundamental internal MT excitation appears when a motor (or motor
complex) acts along a single PF with index $k$ compressing or stretching it.
The motor energy in this case can be written as $E_{V,mot}=F\left(
u_{k}\left(  s_{0}+\left|  \mu\right|  /2\right)  -u_{k}\left(  s_{0}-\left|
\mu\right|  /2\right)  \right)  \approx F\left|  \mu\right|  \int_{0}%
^{L}\delta\left(  s-s_{0}\right)  u_{k}^{\prime}\left(  s\right)  ds$ where
$\left|  \mu\right|  $ is the size of the motor step. Like in the previous
case it is sufficient to keep the energy contribution of the mode $q=1$ as the
others decouple from each other and from the curvature term. Along very
similar line of derivation as in the S-type excitation case we obtain a
solution which is planar and contained in the plane spanned by the vector
$\vec{r}_{k}$ and $\vec{t}.$ The resulting tangent angles for an excitation in
the middle of the MT ($s_{0}=0$) are given by eq \ref{thetaSType} with
$\Phi_{i}=\Phi_{i}^{V}\left(  s\right)  =\frac{s}{\left|  s\right|  }\left(
\vec{e}_{i}\cdot\Delta\vec{r}_{k}\right)  \frac{\gamma_{i}F\left|  \mu\right|
}{2\left(  1+\alpha\right)  \hat{B}}$ which is now $s$ dependent and changes
sign at $s=0.$ Here $\left(  \gamma_{x},\gamma_{y}\right)  =\sin\left(
\pi/N\right)  \left(  1-\cos\left(  2\pi/N\right)  \right)  ^{-1}$ $\left(
\sin\left(  2\pi\left(  k+1/2\right)  /N\right)  ,-\cos\left(  2\pi\left(
k+1/2\right)  /N\right)  \right)  .$ \begin{figure}[ptb]
\includegraphics*[width=8cm]{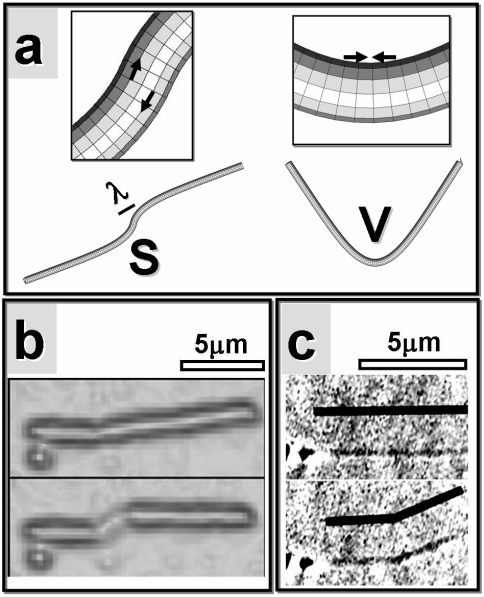}\caption{a: The shapes of V
and S-type excitations as given by Eq. \ref{thetaSType} and their experimental
observations for katanin operating on MTs (b,c). b: Adapted and edge enhanced
image from \cite{MTSTypeKink}. c: Adapted from \cite{MTVKink}, black bars
highlight the MT deformation. The upper and lower images show the MT before
and after kink generation respectively. \ }%
\label{KinkCompilation}%
\end{figure}The resulting shape, that we call a \textit{''V-let''}, is a
smooth V-shaped planar kink in the MT\ backbone with continuous curvature
which relaxes on the length-scale $\lambda.$ While superficially similar the
S-let and V-let solutions are physically very different for two reasons.
First, the $s/\left|  s\right|  $ factor in the V-let solution changes the
symmetry with respect to the S-let which leads to dramatic effects on the
end-end distance as we see below. Second difference lies in the different
scaling of the numerical prefactors which in the case of a V-let do not
contain the screening length $\lambda$ and involve additionally a very large
reduction factor $1/\left(  1+\alpha\right)  .$ In practice this suppresses
significantly the involved deformations ($\theta\approx10^{-5}-10^{-6}$),
orders of magnitude below that of a S-let corresponding to the same force
$(\theta\approx10^{-1})$. However the situation changes dramatically if the
motor is operating at a position of a vacancy in the tubulin lattice, cf. Fig
1. In this case the motor is not hindered by the large rigidity of the short
PF portion that\ the motor acts on. Formally there is no requirement of
continuity for $u_{k}\left(  s\right)  $ of the involved PF $k$ at the
position of the gap. For such a combined defect + motor excitation which we
call $V^{gap}$\textit{-let}, the motor energy is given by $E_{V^{gap}%
,mot}=F\left(  u_{k}\left(  s_{0}+0\right)  -u_{k}\left(  s_{0}-0\right)
\right)  $ and its $U_{x/y}$ dependent component becomes $E_{V^{gap},mot}^{i}$
$=F\gamma_{i}\left(  U_{i}\left(  s_{0}+0\right)  -U_{i}\left(  s-0\right)
\right)  .$ After a short calculation in direct analogy with the previous
cases we obtain the same form as in eq \ref{thetaSType} but with $\Phi
_{i}=\Phi_{i}^{V^{gap}}\left(  s\right)  =\frac{s}{\left|  s\right|  }\left(
\vec{e}_{i}\cdot\Delta\vec{r}_{k}\right)  \frac{\lambda\gamma_{i}F}{\hat{B}}%
$.While its functional form and symmetry coincide with that of the V-let, the
prefactor of a $V^{gap}$-let solution is more similar in magnitude to that of
a S-let. It is intuitive to think of a $V^{gap}$ - let as a V-let with an
effectively renormalized motor force $\tilde{F}=\allowbreak2a^{-1}%
\lambda\left(  1+\alpha\right)  F.$ The distribution of angles of observed
V-shapes measured by Davies et al \cite{MTVKink} of $\theta_{\max}%
\approx15^{\circ}-30^{\circ}$(cf.Fig 2c) would suggest for the defect free
case (V-let) a very large required force of $F=6\times10^{4}-6\times10^{5}pN$.
However , the same estimate for the $V^{gap}$-let case gives much more
moderate $10-100pN$ (few tens of motors), showing the prime importance of
lattice defects in the case of V type excitations. Interestingly Davies et al.
\cite{MTVKink} also suggested a crucial role of lattice defects based on the
pattern and kinetics of MT decomposition by katanin.

\textit{Statistical mechanics of multiple excitations.} It is particularly
interesting to understand the collective contribution of a large number of
internal MT excitations acting at random positions $s_{j}$ and orientations
$\overrightarrow{\mu}_{j}$ in addition to the MT thermal fluctuations. The
motor energy for S- and V-lets becomes now respectively $E_{S,mot}^{i}%
=-\frac{F}{a}\sum\nolimits_{j}\mu_{j}^{i}\hat{\Delta}_{i}\left(  s_{j}\right)
$ and $E_{V,mot}^{i}=-F\sum\nolimits_{j}\left|  \mu\right|  \gamma_{j}%
^{i}U_{i}^{\prime}\left(  s_{j}\right)  $. It is convenient to introduce the
vector $\overrightarrow{\mu}_{S/V,j}$ such that $\mu_{S,j}^{i}\equiv\mu
_{j}^{i}$ and $\mu_{V,j}^{i}\equiv\left|  \mu\right|  \gamma_{j}^{i}$
($i=x,y$).\textbf{\ }For a fixed (but arbitrary) distribution of motors the
partition functions $Z_{i}=\int\emph{D}U_{i}\emph{D}\theta_{i}\exp\left(
-\left(  E_{MT}^{i}+E_{mot}^{i}\right)  /k_{B}T\right)  $ and the correlation
functions $\left\langle \overline{\theta_{i,q}\theta_{i,p}}\right\rangle $
($q,p\neq0$) can be obtained. Here $\left\langle ...\right\rangle $ denotes
the average over the random motore distribution and $\overline{\left(
...\right)  }$ is the average over the thermal noise. $\left\langle
\overline{\theta_{i,q}\theta_{i,p}}\right\rangle $ decomposes into the sum of
a thermal contribution $\overline{\left\langle \theta_{i,q}\theta
_{i,p}\right\rangle }_{T}=2k_{B}TL^{-1}G(q)\delta_{p,q}$ with the propagator
$G(q)=\left(  \hat{B}q^{2}+\frac{q^{2}\hat{K}_{s}a^{2}}{q^{2}+\hat{K}_{s}%
/\hat{K}_{c}}\right)  ^{-1}$ and a motor contribution $\left\langle
\overline{\theta_{i,q}\theta_{i,p}}\right\rangle _{S/V}=\Psi_{S/V}\left(
p\right)  \Psi_{S/V}\left(  q\right)  C_{S/V}\left(  q,p\right)  .$ For S- and
V-let case we have respectively $\Psi_{S}\left(  p\right)  =\frac{2F\hat
{K}_{c}p^{2}}{L\left(  \hat{K}_{c}p^{2}+\hat{K}_{s}\right)  }G(p)$ and
$\Psi_{V}\left(  p\right)  =\frac{2Fa\hat{K}_{s}p}{L\left(  \hat{K}_{c}%
p^{2}+\hat{K}_{s}\right)  }G(p)$, with the motor position and orientation
correlator $C_{S/V}\left(  q,p\right)  =\left\langle \sum\nolimits_{j,l}%
^{N}\mu_{S/V,j}^{i}\mu_{S/V,l}^{i}\cos\left(  qs_{j}\right)  \cos\left(
ps_{l}\right)  \right\rangle $. $C_{S/V}\left(  q,p\right)  $ is easily
computed for the simplest choice of a uniform motor position and orientation
distribution: $P(s_{i})=1/L$ , $\overrightarrow{\mu}_{S/V,j}=\left|  \mu
_{S/V}\right|  \left(  \cos\phi_{j},\sin\phi_{j}\right)  $ with an random
angle $\phi_{j}$ with a probability distribution $P\left(  \phi_{j}\right)
=1/2\pi$. In this case we obtain the length reduction $\left\langle
\overline{\Delta z}\right\rangle /L\approx\tfrac{1}{4}\sum_{q}\left(
\left\langle \overline{\theta_{x,q}^{2}}\right\rangle +\left\langle
\overline{\theta_{y,q}^{2}}\right\rangle \right)  $which naturally decomposes
in a sum of a thermal fluctuation term and a motor term $\left\langle
\overline{\Delta z}\right\rangle =\left\langle \overline{\Delta z}%
\right\rangle _{T}+\left\langle \overline{\Delta z}\right\rangle _{S/V}$. In
the relevant limiting case $L/\lambda\gg1$ we obtain for the thermal part%
\begin{equation}
\frac{\left\langle \overline{\Delta z}\right\rangle _{T}}{L}\approx\frac
{L}{6l_{p}^{\infty}}+\dfrac{1}{2}\sqrt{\frac{k_{B}T}{a^{2}\hat{K}_{s}l_{p}%
^{0}}} \label{ThermalSlack}%
\end{equation}
with the large and small scale persistence lengths given by $l_{p}^{\infty
}=\left(  \hat{B}+a^{2}\hat{K}_{c}\right)  /k_{B}T$ \ and $l_{p}^{0}=\left(
\allowbreak\alpha/\left(  1+\alpha\right)  \right)  ^{-3}\hat{B}/k_{B}T$.
Interestingly the term $a^{2}\hat{K}_{s}$ can be formally understood as an
intrinsic self-tension straightening the MT at small scales\textbf{. }Similar
formulas appear in different geometries for the railway-track\cite{Everaers}
and wormlike-bundle model\cite{Heussinger}. The motor dependent length
reduction for the S-, V- and $V^{gap}$-let excitations with line density
$\rho$ is given by:%

\begin{gather}
\frac{\left\langle \overline{\Delta z}\right\rangle _{S}}{L}=c_{S}\frac{\rho
F^{2}\mu^{2}}{a^{4}\hat{K}_{s}^{2}\lambda}\label{deltazs}\\
\frac{\left\langle \overline{\Delta z}\right\rangle _{V}}{L}=c_{V}\frac{\rho
F^{2}\mu^{2}L}{a^{2}\hat{K}_{s}^{2}\lambda^{4}}\allowbreak\text{ ,\ }%
\frac{\left\langle \overline{\Delta z}\right\rangle _{V^{gap}}}{L}=c_{V^{gap}%
}\frac{\rho F^{2}\mu^{2}L}{a^{4}\hat{K}_{s}^{2}\lambda^{2}} \label{deltazv}%
\end{gather}
with $c_{S}$ $=\alpha^{2}\left(  1+\alpha\right)  ^{-2}/16$ $\approx
6\times10^{-2},$ $c_{V}$ $=0.18\alpha^{2}\left(  1+\alpha\right)  ^{-4}$
$\approx1.3\times10^{-4}$ and $c_{V^{gap}}$ $=0.73\alpha^{2}\left(
1+\alpha\right)  ^{-2}$ $\approx0.7.$ Remarkably the S- and V/$V^{gap}$-lets
show different scaling. In particular $\left\langle \overline{\Delta
z}\right\rangle _{V/V^{gap}}/L$ grows with $L\;$(in analogy to the first term
in the thermal contribution \ref{ThermalSlack}) while $\left\langle
\overline{\Delta z}\right\rangle _{S}/L$ stays length independent. The
physical reason for this difference becomes obvious from Fig 2, as the
relative slack $\left\langle \overline{\Delta z}\right\rangle /L$ induced by a
single S-let scales with $\lambda/L$, while for an V/$V^{gap}$-let it is
essentially length independent. For longer MTs this effect leads to a strong
dominance of $V^{gap}$-lets over S-lets $\left\langle \overline{\Delta
z}\right\rangle _{V^{gap}}/\left\langle \overline{\Delta z}\right\rangle
_{S}\sim\left(  L/\lambda\right)  \gg1$. Although having the same $L$ scaling
the minute prefactor of defect free V-lets renders their contribution
relatively insignificant $\left\langle \overline{\Delta z}\right\rangle
_{V}/\left\langle \overline{\Delta z}\right\rangle _{S}\sim a^{2}\lambda
^{-3}L\approx10^{-6}$ even for very long MTs ($L=100\mu m$), underlining the
importance of lattice vacancies transforming a V-let into a $V^{gap}$-let.
Another interesting observation is that in all three cases $\left\langle
\overline{\Delta z}\right\rangle _{S/V/V^{gap}}\propto\rho F^{2}$. Fixing the
number of motors $N_{mot}$ but regrouping them into $N_{mot}/M$ clusters of
size $M$ we have $\rho\rightarrow M^{-1}\rho$, $F\rightarrow MF$ \ and
therefore $\left\langle \overline{\Delta z}\right\rangle _{S/V/V^{gap}}\propto
M,$ i.e. the slack grows linearly with the cluster size. This indicates that
cooperativity (positional correlation) of motor action can lead to strong
enhancement of the slack length.

From Eqs.\ref{ThermalSlack}-\ref{deltazv} we can derive criteria
for the dominance of motor slack over the thermal slack. For
instance using the values estimated from Fig 2 b,c for katanin for
elastic constants from \cite{Pampaloni}\ ($F=20pN$, $\lambda=1\mu
m$) and $\mu=8nm$, $L=20\mu m$ we obtain $\left\langle
\overline{\Delta z}\right\rangle /L=\rho/\rho_{c}$ with
$\rho_{c,S}\approx0.25nm^{-1},$ \
$\rho_{c,V^{gap}}\approx1.2\times 10^{-3}nm^{-1}.$ For large
enough motor densities the katanin action easily
dominates over the thermal slack $\left\langle \overline{\Delta z}%
\right\rangle _{T}/L\approx6\times10^{-4}$.

Being evolutionary specialized for MT deformation and degradation katanin is
likely to be among the strongest slack generating motors. We suspect however
that classical motors like dynein and kinesin might cause less pronounced but
observable effects as well. While dynein is known to walk between several PFs,
kinesin is very strictly following a single one\cite{Kinesin/Dynein}. Our
theory suggest that dynein should induce moving S-lets, yet with quickly
fluctuating signs which would diminish the effect considerably. A battery of
many kinesins, however, walking over a MT\ region with many tubulin vacancies,
would give rise to spatially stationary $V^{gap}$-lets blinking between ''on''
and ''off'' states. The theoretical and experimental exploration of these
issues is an interesting future direction.

The authors acknowledge fruitful discussions with E. Frey, C. Heussinger,
M.Bathe, O. Campas, J.F. Joanny and P.C. Nelson. I.M.K. acknowledges support
by the Max-Planck Society.


\begin{thebibliography}{99}
\bibitem{Kinesin/Dynein}J. Howard, Mechanics of Motor Proteins and the
Cytoskeleton. Sinauer Press (2001); L.A. Amos and W.G. Amos, Molecules of the
Cytoskeletion, Guilford (1991).

\bibitem {Kis}A. Kis et al. Phys. Rev. Lett. 89: 248101 (2002)

\bibitem {Pampaloni}F. Pampaloni et al. PNAS 103: 10248 (2006)

\bibitem {Everaers}R. Everaers, R. Bundschuh, and K. Kremer, Europhys. Lett
29, 263 , (1995)

\bibitem {Heussinger}C. Heussinger, M. Bathe and E. Frey, [cond-mat/0702097]

\bibitem {MTVKink}L.J. Davis, D.J. Odde, S.M. Block, and S.P. Gross, Biophys.
J. 82, 29162927 (2002)

\bibitem {MTSTypeKink}J.J. Hartmann et. al, Cell, Vol. 93, 277287; Movie at
http://valelab.ucsf.edu/images/mov-rhomtsvkat.mov with kind permission by R. Vale
\end{thebibliography}
\end{document}